# ON-MASS-SHELL RENORMALIZATION OF THE $\langle \bar{t}t \rangle$ CONTRIBUTION TO THE $\Delta I = 1/2\ s \leftrightarrow d$ SELF-ENERGY TRANSITION

M.R. AHMADY

*International Centre for Theoretical Physics, Miramare P.O. Box 586, 34100 Trieste, Italy*

V. ELIAS and N.C.A. HILL

*Department of Applied Mathematics, University of Western Ontario, London, Ontario N6A 5B7, Canada*

## ABSTRACT

In $\langle \bar{t}t \rangle$-scenarios for dynamical electroweak symmetry breaking, the presence of a very large t-quark condensate necessarily generates an $s \leftrightarrow d$ self-energy transition which will contribute exclusively to $\Delta I = 1/2$ strangeness-changing matrix elements. This contribution is calculated and compared to the purely perturbative contribution, which is well-known to be far too small to have an appreciable effect on $\Delta I = 1/2$ strangeness-changing matrix elements after mass-shell renormalization subtractions are taken into account. The ratio between the t-quark condensate's contribution and the standard-model perturbative contribution to the mass-shell renormalized $s \leftrightarrow d$ self-energy transition is shown to be of order $\mid m_t \langle \bar{t}t \rangle \mid /M_W^4$, suggesting that the large t-quark condensate proposed for electroweak symmetry breaking may also be responsible for the $\Delta I = 1/2$ enhancement of strangeness-changing decays.

If electroweak symmetry breaking is driven dynamically by a t-quark condensate $\langle \bar{t}t \rangle$, such a condensate would necessarily have an enormous magnitude. A symmetry-breaking scenario based upon a Nambu-Jona-Lasinio mechanism would suggest a condensate of magnitude $-m_t \Lambda^2$, where $\Lambda$, the NJL cutoff, is certainly larger than $m_t$ and may even correspond to an order-Planck-mass unification mass scale.[1,2] Such a large condensate has already been shown not to alter salient electroweak results, such as the less-than-1% deviation of the electroweak $\rho$-parameter from unity.[1] Nevertheless, it is of genuine importance to examine how such a very large condensate of *standard- model* fermions, as opposed to technifermions outside the purview of CKM mixing, might enter other electroweak results. For example, the presence of a very large t-quark condensate ($\langle \bar{t}t \rangle$) necessarily generates an $s \leftrightarrow d$ self-energy transition (via $\bar{t}s$ and $\bar{s}t$ CKM mixing angles[3]) which will contribute exclusively to $\Delta I = 1/2$ strangeness -changing matrix elements.

Of course, there already exists a purely perturbative $s \leftrightarrow d$ self-energy transition within the standard model, but this contribution is known to be two to three orders of magnitude too small to have an appreciable effect on $\Delta I = 1/2$ strangeness-changing matrix elements after mass-shell renormalization subtractions are taken into account.[4] In unitary gauge, the unrenormalized $s \leftrightarrow d$ self-energy amplitude is of the form

$$\Sigma_{ds}(p) = H(p^2)\slashed{p}(1-\gamma_5), \quad (1)$$

where $H(p^2)$ contains both constant and momentum-dependent terms (the former being divergent). However, the on-mass-shell renormalization conditions $\bar{u}_d(p)[\Sigma_{ds}(p)]^{ren} = 0, [\Sigma_{ds}(p)]^{ren} u_s(p) = 0$, provide sufficient information to determine algebraically the four allowed counterterms contributing to $[\Sigma_{ds}(p)]_{ren} = \Sigma_{ds}(p) + A\slashed{p}(1-\gamma_s) + B\slashed{p}(1+\gamma_s) + C + D\gamma_s$.[5] If we approximate $H(p^2)$ with $H(0) + p^2 H'(0)$, we find that[4]

$$\Sigma_{ds}(p)^{ren} = H'(0)\{[p^2 - m_s^2 - m_d^2]\slashed{p}(1-\gamma_s) + m_s m_d[-\slashed{p}(1+\gamma_5) + (m_s+m_d) - \gamma_s(m_s-m_d)]\}, \quad (2)$$

where the t-quark's Fig. 1 (f1) contribution is given by[4]

$$[H'(0)]_{f1} = (G_F/\sqrt{2})(V_{td})^* V_{ts}(B_t + 5/12)/4\pi^2, \quad (3)$$

with $B_t$ between -1.0 and -1.3 for $m_t$ between 1.5 $M_W$ and 2.0 $M_W$. Using the present empirical range of CKM mixing angles $[0.003 \leq |V_{td}| \leq 0.018, 0.030 \leq |V_{ts}| \leq 0.054]$,[6] we see that $1 \cdot 10^{-11} GeV^{-2} \leq [H'(0)]_{f1} \leq 2 \cdot 10^{-10} GeV^{-2}$. The renormalized transition amplitude (2) can be incorporated into the $\Delta I = 1/2$ component of the $K_s \to 2\pi$ amplitude,

denoted by [4,7]

$$a_{1/2} = 3\sqrt{2}g_{Kqq}(m_s - m_d)M^4(1 - m_\pi^2/m_K^2)H'(0)/(64\pi^2 f_\pi^2) \quad (4)$$

where M is a phenomenological cut-off of order 2 GeV and $g_{Kqq} \approx 3.8$. Using (3) and (4) we find the t-quark's purely perturbative contribution to $a_{1/2}$ to be between $2 \cdot 10^{-9}$ and $9 \cdot 10^{-11}$ GeV, a number at least two orders of magnitude smaller than the experimental value ($3.5 \cdot 10^{-7}$GeV) reflecting the enhancement of the $\Delta I = 1/2$ component of strangeness-changing nonleptonic decays. A more careful treatment[8] of the hadronization of $\Sigma_{ds}(p)^{ren}$, based upon a direct comparison of the weak kaon axial-vector current $-i \int d^4x e^{iq \cdot x} \langle 0|T[A_\mu^3(x)H_W^{\Delta S=1}]|\bar{K}^0\rangle$ to the strong-interaction axial-vector current $i\sqrt{2}f_K q_\mu = -i \int d^4x e^{iq \cdot x} \langle 0|A_\mu^{6+i7}(x)|\bar{K}^0\rangle$, leads to virtually the same conclusions: $[H'(0)]_{f1}$ is between two and four orders of magnitude too small to account for the measured $K \to 2\pi$ decay rate.

If electroweak symmetry breaking is caused by condensation of a t- quark-antiquark pair, residual contributions to the Wick-Dyson expansion of Feynman amplitudes involving the vacuum expectation values of normal ordered t-quark and antiquark fields are no longer seen to vanish. The presence of such contributions is entirely analogous to how $\langle u\bar{u} \rangle$ and $\langle d\bar{d} \rangle$ condensates characterizing the chiral- symmetry breakdown of QCD enter Feynman amplitudes that are utilized in the construction of QCD sum-rules.[8,9] In particular, the unitary gauge contribution of Fig. 2, corresponding to replacing the time ordered vacuum expectation value of interaction-picture t-quark fields $\langle 0|T[t(x)\bar{t}(0)]|0\rangle$ with the corresponding normal-ordered contribution $\langle 0| : t(x)\bar{t}(0) : |0\rangle$ in the Wick-Dyson expansion of $\langle 0|T[t_H(x)\bar{t}_H(0)]|0\rangle$ [the fermion two-point function of Heisenberg fields], is seen to be

$$\begin{aligned}\Sigma_{ds}(p) &= \frac{G_F M_W^2}{\sqrt{2}} V_{td}^* V_{ts} \int \frac{d^4q}{(2\pi)^4} \int d^4x e^{i(p-q)\cdot x} \\ &\times \left[\frac{g^{\mu\nu} - q^\mu q^\nu/M_W^2}{q^2 - M_W^2}\right] \gamma_\mu (1-\gamma_5)\langle 0| : t(x)\bar{t}(0) : |0\rangle \gamma_\nu (1-\gamma_5).\end{aligned} \quad (5)$$

The $\langle \bar{t}t \rangle$ projection of $\langle 0| : t(x)\bar{0}(x) : |0\rangle$ is given by[10]

$$\langle 0| : t(x)\bar{t}(0) : |0\rangle = -\frac{\langle \bar{t}t \rangle}{6m_t^2}(i\gamma \cdot \partial + m_t)[J_1(m_t\sqrt{x^2})/\sqrt{x^2}]. \quad (6)$$

The theoretical consistency of eq. (6) has been examined in a number of places. Insertion of this vacuum expectation value into flavour- diagonal fermion two-point functions has been shown to uphold the gauge invariance of the fermion propagator pole in both covariant [11] and noncovariant [12] gauges, despite strong overall gauge dependence of the fermion self-energy. Moreover, inclusion of the $\langle \bar{f}f \rangle$ projection of the $\langle 0| : f(x)\bar{f}(y) : |0\rangle$ contribution to the Wick-Dyson expansion of the vacuum polarization of a vector-coupled spin-1 field has been shown to yield only transverse contributions, consistent with gauge invariance and the vector- current Ward identity.[13] Similar inclusion of the $\langle \bar{f}f \rangle$ projection of the $\langle 0| : f(x)\bar{f}(y) : |0\rangle$-contributions to the Wick-Dyson expansion of the VVA-triangle amplitude has also been demonstrated to uphold the anomalous axial-vector Ward identity.[14]

Equation (6) may be re-expressed as follows:[15]

$$\langle 0| : t(x)\bar{t}(0) : |0\rangle \equiv \int d^4k e^{-ik\cdot x}(\not{k}+m_t)F_t(k), \tag{7}$$

$$\int d^4k e^{-ik\cdot x}F_t(k) = -\langle \bar{t}t \rangle J_1(m_t\sqrt{x^2})/(6m_t^2\sqrt{x^2}), \tag{8}$$

$$\int d^4k e^{-ik\cdot x}g(k^2)F_t(k) = \int d^4k e^{-ik\cdot x}g(m_t^2)F_t(k), \tag{9}$$

where the last property [eq. (9)] is a consequence of $\langle 0| : t(x)\bar{t}(0) : |0\rangle$ being a solution of the free-field Dirac equation. One finds from Eq.(5) that the coefficient $H(p^2)$ in the unrenormalized Fig. 2 (f2) self-energy (1) is given by

$$[H(p^2)]_{f2} = \frac{G_F}{\sqrt{2}} M_W^2 V_{td}^* V_{ts} \left\{ \left(-4 - \frac{2(m_t^2+p^2)}{M_W^2}\right) \left[\frac{\langle \bar{t}t \rangle}{24m_t p^2}\right.\right.$$
$$\left.\left. + \frac{p^2+m_t^2-M_W^2}{2p^2}I(p^2)\right] + \frac{4m_t^2}{M_W^2}I(p^2)\right\}, \tag{10}$$

$$I(p^2) = \int d^4k F_t(k)/[(p-k)^2 - M_W^2 + i\epsilon]. \tag{11}$$

The integral in Eq. (11) may be evaluated first by exponentiating the propagator and through utilization of Eqs. (9) and (8):[15]

$$\begin{aligned}I(p^2) &= -i\int_0^\infty d\eta e^{i\eta(p^2+m_t^2-M_W^2+i\epsilon)} \int d^4k e^{-i(2\eta p)\cdot k} F_t(k) \\ &= \frac{i\langle \bar{t}t\rangle}{6m_t^2} \int_0^\infty d\eta\, exp[i\eta(p^2+m_t^2-M_W^2+i\epsilon)]\frac{J_1(2m_t\eta\sqrt{p^2})}{2\eta\sqrt{p^2}} \\ &= -\frac{\langle \bar{t}t\rangle}{24m_t^3 p^2}\left[(p^2+m_t^2-M_W^2) - \sqrt{(p^2-m_t^2)^2 - M_W^2(2p^2+2m_t^2-M_W^2)}\right],\end{aligned} \tag{12}$$

where the final line of Eq.(12) is obtained by use of a tabulated integral.[16] Substitution of the final line of (12) into (10) reveals that the fig. 2 contribution to $H(p^2)$ is analytic at $p^2 = 0$:

$$[H(p^2)]_{f2} = [H(0)]_{f2} + p^2[H'(0)]_{f2}, \tag{13}$$

$$[H(0)]_{f2} = -\frac{G_F V_{td}^* V_{ts} \langle \bar{t}t \rangle}{4\sqrt{2}(m_t^2 - M_W^2)^2}[1 - 7M_W^2/3m_t^2 - 2M_W^4/3m_t^4], \qquad (14)$$

$$[H'(0)]_{f2} = \frac{G_F V_{td}^* V_{ts} \langle \bar{t}t \rangle m_t^5}{12\sqrt{2}(m_t^2 - M_W^2)^4}[1 - 4M_W^2/m_t^2 + 9M_W^4/m_t^4]. \qquad (15)$$

The relative contributions of Fig. 2 and Fig. 1 to the on-shell- renormalized self-energy transition, as defined by (2), may be obtained via direct comparison of Eqs. (3) and (15). One finds for $m_t$ between 120 GeV and 160 GeV that $|12m_t\langle\bar{t}t\rangle/M_W^4| \geq |[H'(0)]_{f2}/[H'(0)]_{f1}| \geq |0.4m_t\langle\bar{t}t\rangle/M_W^4|$. As remarked earlier, the Fig. 1 contribution $[H'(0)]_{f1}$ to the $\Delta I = 1/2$ component of the $K \to 2\pi$ decay amplitude is between two and four orders of magnitude below the experimental value. Consequently, we see that the Fig. 2 contribution $[H'(0)]_{f2}$ yields a contribution *comparable to the experimental ampitude* provided $\langle\bar{t}t\rangle \sim m_t^3$ for maximal mixing angles and $m_t \approx 120 GeV$. For minimal mixing angles and $m_t \approx 160 GeV$, the value $\langle\bar{t}t\rangle \sim (10m_t)^3$ is compatible with the experimental $K \to 2\pi$ decay rate.

These results suggest, within the context of t-quark-condensate driven electroweak symmetry breaking, that the large magnitude of the t-quark condensate may be responsible for the $\Delta I = 1/2$ rule. However, these results appear to rule out any sensitivity of the t-quark condensate to Planck-mass unification scales.

**Figure Captions**

**Figure 1:** Purely perturbative t-quark contribution to the s-d transition amplitude.

**Figure 2:** $\langle \bar{t}t \rangle$ contribution to the s-d transition amplitude.